\documentclass[%
 aps, prd, twocolumn, nofootinbib, superscriptaddress,10pt]{revtex4-2}
\usepackage{amsmath,amssymb,bm}
\usepackage{graphicx}
\usepackage{physics}
\usepackage{comment}
\usepackage[colorlinks=true,linkcolor=blue,citecolor=blue,urlcolor=blue]{hyperref}
\begin{document}
\title{Singular Asymptotics of SPADE in Quantum Source Discrimination}
\author{Natsuki Kariya}
\email{natsuki.kariya@mizuho-rt.co.jp}
\affiliation{Mizuho Bank, Informatics and Mathematics Research Institute, 1--5--5 Otemachi, Chiyoda-ku, Tokyo 100-8176, Japan}
\date{\today}

\begin{abstract}
We study far-field discrimination between one and two incoherent point sources in the singular regime of weak and closely spaced emitters. Under ideal alignment, spatial-mode demultiplexing (SPADE) attains the quantum-optimal large-sample Stein exponent, but the finite-photon behavior near the one-source boundary and the effect of realistic imperfections remain less understood. Using singular learning theory, we analyze both the aligned and misaligned problems. In the aligned Gaussian case, for prior densities that are smooth and strictly positive in the physical $(\epsilon,s)$ coordinates near the singularity of the aligned model, direct imaging and SPADE share the same real log canonical threshold $\lambda=1/2$ but have different multiplicities, yielding distinct Bayes free-energy asymptotics. A fixed nonzero misalignment removes the exact support mismatch of ideal SPADE: locally, the fixed-offset binary-SPADE Kullback--Leibler function has the normal-crossing form $K\asymp\epsilon^2s^2$, giving $(\lambda,m)=(1/2,2)$ under the same prior class, as for direct imaging. The local separation scale nevertheless depends on the source-position convention. Moreover, full Hermite--Gaussian mode counting about an offset sorter axis has a reflection-induced KL-zero branch at $s^\ast=2\theta$, where the two-source alternative becomes indistinguishable from the null. Pointwise finite-$n$ binary-SPADE calculations exhibit the corresponding power collapse. These results identify measurement support, nuisance geometry, prior weighting, and identifiability as structural ingredients that must be tracked in finite-photon quantum discrimination.

\end{abstract}

\maketitle
\section{Introduction}

Determining whether observed light originates from one incoherent point source or from two is a minimal but fundamental problem in optical detection. Despite its apparent simplicity, it already captures the essential difficulty of resolving faint and closely spaced emitters, a situation that arises naturally in microscopy and observational astronomy. The problem is especially challenging in the practically important regime where the source separation is very small or one source is much weaker than the other.

The Rayleigh criterion has traditionally been regarded as the classical resolution limit for resolving two closely spaced incoherent sources by direct imaging. More recently, however, spatial-mode demultiplexing (SPADE) was proposed as a quantum-measurement-inspired detection scheme and shown to overcome Rayleigh's curse and, under ideal conditions, to attain the quantum-optimal limit for source discrimination and separation~\cite{Tsang2016,NairTsang2016,LuKroviNairGuhaShapiro2018}.

Experimental demonstrations and analyses of more realistic settings soon followed, including studies of source asymmetry, misalignment, and imperfect demultiplexing~\cite{Paur2016,Almeida2021,Zhang2024UnequalBrightness}. Robustness to an unknown centroid has also motivated adaptive and hybrid schemes in which the centroid is estimated before mode sorting; this direction has been further developed in subsequent theoretical and experimental work~\cite{Tsang2016,Grace2020,OzerGrace2024}. In parallel, Schlichtholz et al.~\cite{Schlichtholz2024} showed that quantum-optimal tests constructed for aligned SPADE can lose discrimination power under demultiplexer misalignment, sharply illustrating the sensitivity of the aligned benchmark.

The difficulty of source discrimination in these regimes has a precise statistical origin. When the two sources are extremely close, or when the secondary source is extremely weak, different parameter values of the two-source model can generate the same observable probability distribution, including the distribution of the one-source null model. The parameterization then ceases to distinguish observable distributions one by one. Equivalently, the map from model parameters to probability distributions is no longer one-to-one, and the Fisher information matrix becomes singular. In statistics, such models lie outside the standard regular framework; consequently, familiar results such as Fisher-information-based asymptotics and the standard $\chi^2$ asymptotics for likelihood-ratio tests need not hold~\cite{Hartigan1985,ChenLi2009}. Although singular models are ubiquitous in modern statistics, a general framework for their analysis---singular learning theory---has emerged from Bayesian statistics only relatively recently and is still evolving~\cite{Watanabe2009,KariyaWatanabe2022,KariyaWatanabe2020}. The present one-versus-two-source discrimination problem is a natural setting in which these ideas become physically relevant, because the one-source distribution is represented non-identifiably along a KL-zero subset of the two-source parameter space.

A central distinction in this paper is between two different operational questions. The usual Stein benchmark asks how rapidly the type-II error decays for a specified two-source alternative with fixed $(\epsilon,s)$. By contrast, when the two-source model is equipped with a prior over unknown source parameters, the marginal law asks for evidence for the two-source model as a whole. In that prior-predictive problem, the marginal likelihood becomes increasingly localized near the Kullback--Leibler zero set as $n$ grows, so alternatives arbitrarily close to the one-source boundary control the asymptotics. Singular learning theory quantifies how this localization can turn a fixed-parameter exponential comparison into a polynomial model-evidence problem.

In this paper, we develop a singular-learning-theoretic analysis of far-field one-versus-two-source discrimination and supplement it with finite-$n$ calculations under a fixed residual misalignment. We first derive the aligned zeta-function structure and validate the resulting finite-$n$ Bayes free-energy approximation over bounded prior windows. We then analyze how a fixed offset changes the local Kullback--Leibler geometry and examine its pointwise Neyman--Pearson consequences.

\medskip
\noindent\textbf{Contributions and scope.}
First, in the aligned Gaussian model, for a prior density that is smooth and strictly positive in the physical $(\epsilon,s)$ coordinates near the singularity of the aligned model, direct imaging and SPADE have the same real log canonical threshold, $\lambda=1/2$, but different multiplicities. This yields a logarithmic subleading distinction within that prior class; the dependence of the pole data on the local prior measure is analyzed in Appendix~E.

Second, for any fixed nonzero sorter offset, the exact support mismatch responsible for the aligned SPADE behavior is removed. In the fixed-primary binary-SPADE model the local Kullback--Leibler function becomes $K\asymp\epsilon^2s^2$, so under the same smooth-positive prior class its pole data are $(\lambda,m)=(1/2,2)$, matching direct imaging. The local separation scale still depends on the nuisance direction: it is $s=O(n^{-1/2})$ for fixed $\epsilon$ in the fixed-primary convention and $s=O(n^{-1/4})$ for generic fixed offsets in a centroid-preserving convention.

Third, the fixed-offset Hermite--Gaussian mode-counting law has a reflection symmetry about the sorter axis. Consequently, the global KL-zero set contains an additional branch $s^\ast=2\theta$, at which even the full mode-resolved SPADE photon-counting distribution of the alternative coincides with the null. The binary-SPADE finite-$n$ stress test makes the resulting loss of identifiability explicit. The purpose of this calculation is to isolate the singular structure created by a residual offset; adaptive recentering remains a distinct receiver-design problem.

\section{Problem setup and large-sample benchmarks}

We briefly recall the large-$n$ benchmarks for one-versus-two-source discrimination that will be needed below, following Ref.~\cite{HuangLupo2021}, and fix notation.
We consider a one-dimensional diffraction-limited imaging system with normalized point-spread function (PSF) $\psi(x)$ satisfying $\int dx\,|\psi(x)|^2=1$. Since the absolute source position is irrelevant for the present discussion, we place the bright source at the origin. The null hypothesis $H_0$ corresponds to a single incoherent source, whereas under the alternative hypothesis $H_1$ a second incoherent source of relative brightness $\epsilon \ll 1$ is present at signed displacement $s$ from the primary source (with physical separation $|s|$):

\begin{align}
p_0(x) &= |\psi(x)|^2, \\
p_1(x) &= (1-\epsilon)|\psi(x)|^2+\epsilon |\psi(x-s)|^2 .
\label{eq:p0p1}
\end{align}

We focus on the sub-Rayleigh weak-companion regime, $\epsilon \ll 1$ and $|s|$ smaller than the width of the PSF.
In direct imaging, each detection event returns a position $X_i$ on the one-dimensional image plane, so that after $n$ detections one obtains independent samples $X^n=(X_1,\dots,X_n)$ from either $p_0$ or $p_1$. For a fixed type-I error tolerance $0<\alpha<1$, let $\beta_n^{\mathrm{DI}}(\alpha)$ denote the minimum type-II error achievable by direct imaging after $n$ detected photons. Classical Stein's lemma gives

\begin{equation}
-\lim_{n\to\infty}\frac{1}{n}\log \beta_n^{\mathrm{DI}}(\alpha)=D(p_0\|p_1),
\label{eq:classicalStein}
\end{equation}

where
\[
D(p_0\|p_1):=\int dx\, p_0(x)\log\frac{p_0(x)}{p_1(x)} .
\]

For a Gaussian PSF,
\[
\psi(x)=\left(\frac{1}{2\pi\sigma^2}\right)^{1/4}
\exp\!\left(-\frac{x^2}{4\sigma^2}\right),
\]

the weak-companion expansion yields

\begin{equation}
D(p_0\|p_1)=\frac{\epsilon^2 s^2}{2\sigma^2}
+O\!\left(\epsilon^2 s^4+\epsilon^3 s^4\right).
\label{eq:classicalKL}
\end{equation}

Thus, for direct imaging, the Stein exponent vanishes as $\epsilon^2 s^2$ in the limit of a faint and nearby secondary source.

A quantum description associates to a single detected photon the states

\[
|\psi_{x}\rangle
:=
\int dx'\,\psi(x'-x)\,a^\dagger(x')|0\rangle,
\]

so that

\begin{align}
\rho_0 &= |\psi_{0}\rangle\langle \psi_{0}|, \\
\rho_1 &= (1-\epsilon)|\psi_{0}\rangle\langle \psi_{0}|
+\epsilon |\psi_{s}\rangle\langle \psi_{s}|.
\label{eq:rho0rho1}
\end{align}

Let $\beta_n^{\mathrm{Q}}(\alpha)$ denote the minimum type-II error optimized over all quantum measurements. Quantum Stein's lemma then implies
\begin{equation}
-\lim_{n\to\infty}\frac{1}{n}\log \beta_n^{\mathrm{Q}}(\alpha)
= D(\rho_0\|\rho_1),
\label{eq:quantumStein}
\end{equation}
where
\[
D(\rho\|\sigma):=\Tr\!\left[\rho(\log \rho-\log \sigma)\right].
\]

For the rank-two model in Eq.~\eqref{eq:rho0rho1}, one has
\[
D(\rho_0\|\rho_1)=\bigl(1-|\langle \psi_0|\psi_s\rangle|^2\bigr)\epsilon+O\!\left(\epsilon^2|\log\epsilon|\right).
\]

For the Gaussian PSF, this becomes
\begin{equation}
D(\rho_0\|\rho_1)=\frac{\epsilon s^2}{4\sigma^2}
+O\!\left(\epsilon s^4+\epsilon^2|\log\epsilon|\right).
\label{eq:quantumKL}
\end{equation}

The quantum benchmark therefore improves the scaling of the error exponent from $\epsilon^2 s^2$ to $\epsilon s^2$.

A concrete measurement that attains this quantum benchmark, under ideal alignment, is spatial-mode demultiplexing (SPADE)~\cite{Tsang2016,HuangLupo2021}. In SPADE, one does not record the continuous image-plane position of each detected photon. Instead, one projects the optical field onto an orthonormal Hermite--Gaussian mode basis $\{\ket{\phi_q}\}_{q=0}^\infty$ adapted to the Gaussian PSF, and records the detected mode label $q$. The corresponding outcome distributions under the null and alternative hypotheses are therefore
\begin{align}
P_0(q)
&=
\Tr\!\left(\Pi_q \rho_0\right)
=
|\langle \phi_q|\psi_0\rangle|^2,
\label{eq:P0SPADE}
\\
P_1(q)
&=
\Tr\!\left(\Pi_q \rho_1\right)
\nonumber\\
&=
(1-\epsilon)|\langle \phi_q|\psi_0\rangle|^2
+
\epsilon |\langle \phi_q|\psi_s\rangle|^2,
\label{eq:P1SPADE}
\end{align}

where $\Pi_q:=\ket{\phi_q}\bra{\phi_q}$. Once the measurement is fixed, SPADE reduces the quantum discrimination problem to an ordinary classical hypothesis test for the discrete outcome distribution $\{P_a(q)\}_{q\ge 0}$. In the perfectly aligned Gaussian setting, the resulting Kullback--Leibler information coincides, to leading order, with the quantum benchmark in Eq.~\eqref{eq:quantumKL}.
The results recalled in this section concern the ideal large-sample regime in which $(\epsilon,s)$ are fixed as $n\to\infty$. They provide the standard reference point for comparing direct imaging, the quantum optimum, and aligned SPADE. However, they do not describe the finite-photon corrections associated with composite alternatives whose marginal likelihood is dominated by neighborhoods of the singular one-source boundary. Those are the regimes addressed in the following sections.

\section{Testing quantities, Bayes factors, and singular asymptotic regimes}

To analyze the finite-photon behavior of source discrimination, we formulate the problem within a Bayesian hypothesis-testing framework that covers both direct imaging and SPADE. The data consist of $n$ independent measurement outcomes
\[
Y^n=(Y_1,\dots,Y_n),
\]
obtained from a fixed measurement. For direct imaging, $Y_i=X_i\in\mathbb{R}$ denotes the detected image-plane coordinate, whereas for SPADE, $Y_i=q_i\in\{0,1,2,\dots\}$ denotes the detected Hermite--Gaussian mode label.

Throughout this paper, the null hypothesis $H_0$ is the one-source model introduced in Sec.~II, while the alternative hypothesis $H_1$ is the two-source model parameterized by
\[
w\in W,
\]
where $w$ collects the relevant parameters, such as the relative brightness $\epsilon$, the source separation $s$. We equip $H_1$ with a prior density $\phi(w)$ on $W$. In the misaligned analysis below, the detector offset $\theta$ is treated as a fixed external parameter rather than as a parameter endowed with a prior, since our interest is in the local behavior with respect to $(\epsilon,s)$ at a given misalignment.
Let $p_0(y)$ denote the distribution of a single measurement outcome under $H_0$, and let $p(y|w)$ denote the corresponding distribution under $H_1$. For direct imaging, these are the image-plane densities in Eqs.~(1) and (2); for SPADE, they are the mode-resolved probabilities in Eqs.~(9) and (10). The corresponding marginal likelihoods are
\begin{align}
m_0(Y^n) &= \prod_{i=1}^n p_0(Y_i), \label{eq:m0-def}\\
m_1(Y^n) &= \int_W \prod_{i=1}^n p(Y_i|w)\,\phi(w)\,dw. \label{eq:m1-def}
\end{align}

It is useful to distinguish two levels of testing quantities.

First, in the large-sample benchmarks of Sec.~II, the type-II errors
\[
\beta_n^{\mu}(\alpha), \qquad \mu\in\{\mathrm{DI},Q,\mathrm{SPADE}\},
\]

refer to ordinary fixed-significance Neyman--Pearson testing at fixed model parameters, such as fixed $(\epsilon,s)$.

Second, once the composite alternative is equipped with the prior $\phi$, it induces the Bayes-averaged alternative law $m_1$ in Eq.~\eqref{eq:m1-def}. The comparison between $m_0$ and $m_1$ is then an ordinary simple-versus-simple testing problem, and its natural likelihood-ratio statistic is the Bayes factor
\begin{equation}
\Lambda_n(Y^n):=\frac{m_1(Y^n)}{m_0(Y^n)}.
\label{eq:Bayes-factor}
\end{equation}

By the Neyman--Pearson lemma, for a prescribed type-I error tolerance $\alpha$, the most powerful test for $m_0$ versus $m_1$ rejects $H_0$ for sufficiently large $\Lambda_n$.

Equivalently, one may use

\begin{equation}
F_n(Y^n):=-\log \Lambda_n(Y^n)
\label{eq:Fn-def}
\end{equation}

as the test statistic, so that the rejection region is defined by $F_n\le \eta$ for an appropriate threshold $\eta$. It is also convenient to introduce the partial free energies

\begin{equation}
F_{0,n}(Y^n):=-\log m_0(Y^n),\qquad
F_{1,n}(Y^n):=-\log m_1(Y^n),
\label{eq:partial-free-energies}
\end{equation}

for which

\begin{equation}
F_n(Y^n)=F_{1,n}(Y^n)-F_{0,n}(Y^n).
\label{eq:free-energy-difference}
\end{equation}

Thus the log Bayes factor is the difference of two free energies in the usual singular-learning-theoretic sense.

The two testing objects answer different questions. Fixed-parameter Neyman--Pearson testing asks how well a specified alternative $(\epsilon,s)$ can be distinguished from the null. The prior-mixture problem instead asks for evidence for the two-source model when its parameters are unknown and represented by a prior predictive distribution. Because that alternative model contains distributions arbitrarily close to the one-source null, its marginal likelihood is asymptotically dominated by a shrinking neighborhood of the KL-zero set. This is why the polynomial singular asymptotics derived below should not be interpreted as a weakening of the fixed-parameter Stein exponent; they quantify a different, model-level discrimination problem.

The central issue is that, in the source-discrimination problem, the null distribution is represented non-identifiably by a subset of the alternative parameter space rather than by an isolated parameter point. To characterize this structure, let
\begin{equation}
K(w):=\int p_0(y)\log\frac{p_0(y)}{p(y|w)}\,d\nu(y)
\label{eq:K-def}
\end{equation}
be the Kullback--Leibler function of the alternative model relative to $H_0$, and define its zero set
\begin{equation}
W_0:=\{w\in W:K(w)=0\}.
\label{eq:KL-zero-set-general}
\end{equation}
In the aligned source model, the null distribution is recovered whenever $\epsilon=0$ or $s=0$, so locally
\begin{equation}
W_0^{\mathrm{al}}=\{\epsilon=0\}\cup\{s=0\}.
\label{eq:aligned-zero-set-general}
\end{equation}
The point $(\epsilon,s)=(0,0)$ belongs to both components of $W_0^{\mathrm{al}}$. The local zeta calculation below is performed in a neighborhood of this point, where both vanishing factors are simultaneously relevant; however, $(0,0)$ is not the unique parameter representation of the one-source distribution. The associated zeta function is
\begin{equation}
\zeta(z):=\int_W K(w)^z \phi(w)\,dw.
\label{eq:zeta-def}
\end{equation}

Here $d\nu(y)$ denotes the underlying measure on the outcome space: it is the Lebesgue measure for direct imaging and the counting measure for SPADE.

In singular learning theory, the asymptotic behavior of the marginal likelihood is governed by the largest pole of $\zeta(z)$ (the minimal review of singular learning theory is given in Appendix D). Denoting by $\lambda$ the real log canonical threshold and by $m$ its multiplicity, one obtains the singular asymptotic expansion
\begin{align}
-\log m_1(Y^n)
&=
-\sum_{i=1}^n \log p_0(Y_i)
+\lambda\log n \nonumber\\
&\quad -(m-1)\log\log n +O_p(1),
\label{eq:singular-marginal}
\end{align}

where the $O_p(1)$ term remains random in the large-$n$ limit. Accordingly, the log Bayes factor, or equivalently the free-energy difference, takes the form

\begin{equation}
F_n(Y^n)=\lambda\log n-(m-1)\log\log n+O_p(1).
\label{eq:singular-Fn}
\end{equation}

It is useful to remark on the relation to the usual KL/Stein regime. The singular expansion in Eq.~\eqref{eq:singular-Fn} is the boundary case relevant to the present source-discrimination problem, in which the alternative family contains the null distribution along the KL-zero set $W_0$ and hence
\[
K_*:=\inf_{w\in W} K(w)=0.
\]
More generally, if the null distribution is not exactly realized within the alternative model, then the leading term of the marginal likelihood is governed by the positive minimum
\[
K_*=\inf_{w\in W}K(w)>0,
\]
and one expects
\[
F_n(Y^n)
=
nK_*+\lambda\log n-(m-1)\log\log n+O_p(1).
\]

In that case, the normalized log Bayes factor satisfies
\[
\frac{1}{n}F_n(Y^n)\to K_* (n\rightarrow \infty),
\]

so the leading behavior is controlled by the minimum Kullback--Leibler information, in direct analogy with the usual large-sample error exponent in Stein-type asymptotics. The present singular regime corresponds to the special boundary situation $K_*=0$, where the $O(\log n)$ terms become leading and reveal the higher-order structure invisible in the ordinary regular picture.

We finally specify the operational testing quantity associated with the prior mixture. For a fixed significance level $0<\alpha<1$, let $T_n(Y^n)\in[0,1]$ denote the conditional probability of rejecting $H_0$, and define the optimal prior-averaged type-II error by
\[
\bar\beta_{n,\phi}^{\mu,*}(\alpha)
:=
\inf_{\substack{0\le T_n\le1\\
\mathbb E_{0}^{\mu}[T_n]\le\alpha}}
\mathbb E_{1,\phi}^{\mu}[1-T_n].
\]
where $\mathbb E_{0}^{\mu}$ and $\mathbb E_{1,\phi}^{\mu}$ are expectations under the null marginal law and the prior-mixture alternative law for measurement $\mu$, respectively. This is the Neyman--Pearson type-II error for the simple marginal laws $m_0$ and $m_1$, rather than an average of parameterwise optimized tests.

A standard change-of-measure argument gives the following implication. If, under the null marginal law,
\[
F_n=a_n+O_p(1),
\]
then, for every fixed $0<\alpha<1$,
\[
-\log \bar\beta_{n,\phi}^{\mu,*}(\alpha)=a_n+O(1).
\]
To see this explicitly, fix $0<\delta<\min\{\alpha,1-\alpha\}$. Tightness of $F_n-a_n$ implies that there is a constant $M$ such that, for all sufficiently large $n$, the typical set
\[
A_{n,M}:=\{|F_n-a_n|\le M\}
\]
has null probability at least $1-\delta$. Since $e^{-F_n}=m_1/m_0$, any level-$\alpha$ test satisfies
\begin{align}
\mathbb E_{1,\phi}^{\mu}[1-T_n]
&=\mathbb E_0^{\mu}\!\left[e^{-F_n}(1-T_n)\right]\nonumber\\
&\ge e^{-a_n-M}\,\mathbb E_0^{\mu}\!\left[(1-T_n)\mathbf 1_{A_{n,M}}\right]\nonumber\\
&\ge e^{-a_n-M}(1-\alpha-\delta).
\label{eq:change-measure-lower}
\end{align}
For the converse bound, $\mathbb P_0^{\mu}(A_{n,M})\ge1-\delta>1-\alpha$, so, allowing boundary randomization when necessary, one can choose a level-$\alpha$ test whose acceptance rule is supported entirely inside $A_{n,M}$ and has null acceptance probability $1-\alpha$. For that test,
\begin{equation}
\mathbb E_{1,\phi}^{\mu}[1-T_n]
\le e^{-a_n+M}(1-\alpha).
\label{eq:change-measure-upper}
\end{equation}
The optimal type-II error is therefore bounded above and below by $n$-independent positive constants times $e^{-a_n}$, which gives the stated $a_n+O(1)$ relation after taking minus logarithms. In the ideal aligned-SPADE setting the null law is degenerate, so exact level $\alpha$ requires randomization on the all-$q=0$ outcome, but the same argument reduces there to an exact constant-factor relation.

This distinction is crucial in the present problem. The deterministic logarithmic terms determine the asymptotic rate of the prior-mixture error, while the $O_p(1)$ remainder describes bounded null fluctuations relevant to finite-$n$ threshold calibration. It is precisely this singular asymptotic structure that we analyze below for the aligned and misaligned source-discrimination problems.

\section{Main results}

We now present the main consequences of the singular-learning-theoretic framework for far-field one-versus-two-source discrimination. Our analysis addresses two complementary questions. First, in the perfectly aligned setting, where SPADE attains the quantum-optimal Stein exponent in the ideal large-sample limit, we ask how model singularity modifies the corresponding Bayes-integrated behavior. Second, in the misaligned setting, the exact support mismatch of ideal alignment is removed, and we examine how this changes the local Kullback--Leibler geometry and the resulting finite-$n$ discrimination problem.

\subsection{Aligned case: singular corrections to Bayes-integrated local behavior}
We first reconsider the perfectly aligned setting from the viewpoint of singular learning theory. To make contact with Sec.~II, we use the same parameters $\epsilon$ and $s$ throughout. The one-source distribution is represented non-identifiably within the two-source family: it is recovered whenever $\epsilon=0$ or $s=0$. Hence the aligned KL-zero set is
\begin{equation}
W_0^{\mathrm{al}}=\{\epsilon=0\}\cup\{s=0\}.
\label{eq:aligned-zero-set-main}
\end{equation}
The point $(\epsilon,s)=(0,0)$ belongs to both components of $W_0^{\mathrm{al}}$. The local zeta analysis below is carried out in a neighborhood of this point, where both vanishing factors are simultaneously relevant; $(0,0)$ should not be confused with the full set of parameter values representing the one-source distribution.
For direct imaging, the relevant Kullback--Leibler information is precisely the divergence introduced in Sec.~II:
\begin{equation}
D_{\mathrm{DI}}(\epsilon,s):=D(p_0\|p_1).
\label{eq:DDI-def}
\end{equation}
For the Gaussian PSF,
\begin{equation}
D_{\mathrm{DI}}(\epsilon,s)
=
\frac{\epsilon^2 s^2}{2\sigma^2}
+O\!\left(\epsilon^2s^4+\epsilon^3s^4\right).
\label{eq:DDI-expansion}
\end{equation}
Equivalently, locally one may write $D_{\mathrm{DI}}(\epsilon,s)=\epsilon^2s^2h_{\mathrm{DI}}(\epsilon,s)$ with $h_{\mathrm{DI}}$ smooth and $h_{\mathrm{DI}}(0,0)=1/(2\sigma^2)>0$.

For aligned SPADE, the null distribution is concentrated on $q=0$, and the Kullback--Leibler divergence between the mode-resolved outcome distributions is therefore exactly
\begin{align}
D_{\mathrm{SPADE}}(\epsilon,s)
&:=
\sum_{q=0}^\infty P_0(q)\log\frac{P_0(q)}{P_1(q)}\nonumber\\
&=
-\log\!\left[1-\epsilon\left(1-e^{-\tau}\right)\right],
\qquad \tau:=\frac{s^2}{4\sigma^2}.
\label{eq:DSPADE-def}
\end{align}
Hence
\begin{equation}
D_{\mathrm{SPADE}}(\epsilon,s)
=
\frac{\epsilon s^2}{4\sigma^2}
+O\!\left(\epsilon s^4+\epsilon^2s^4\right).
\label{eq:DSPADE-expansion}
\end{equation}
Thus $D_{\mathrm{SPADE}}(\epsilon,s)=\epsilon s^2h_{\mathrm{SPADE}}(\epsilon,s)$ with a smooth positive local factor satisfying $h_{\mathrm{SPADE}}(0,0)=1/(4\sigma^2)$. This leading term agrees with the quantum Stein benchmark in Eq.~\eqref{eq:quantumKL}.

The singular-learning-theoretic quantities governing the local subleading asymptotics are obtained from the zeta function associated with the Kullback--Leibler information. For each scheme $\mu\in\{\mathrm{DI},\mathrm{SPADE}\}$, define
\begin{equation}
\zeta_\mu(z):=\int D_\mu(\epsilon,s)^z \,\phi(\epsilon,s)\,d\epsilon\,ds,
\label{eq:zeta-mu}
\end{equation}
where $\phi(\epsilon,s)$ is a prior density that is smooth and strictly positive in a neighborhood of $(\epsilon,s)=(0,0)$, with the aligned KL-zero set given by Eq.~\eqref{eq:aligned-zero-set-main}. This assumption specifies the prior class for the result below. In singular learning theory the pole data belong to the local Kullback--Leibler function together with the prior measure; Appendix~E makes this dependence explicit.

After restricting the integration to a sufficiently small neighborhood of $(\epsilon,s)=(0,0)$, the smooth positive factors can be absorbed into local constants and one finds
\begin{equation}
\zeta_{\mathrm{DI}}(z)=\frac{C_{\mathrm{DI}}}{(2z+1)^2},
\qquad
\zeta_{\mathrm{SPADE}}(z)=\frac{C_{\mathrm{SPADE}}}{(z+1)(2z+1)},
\label{eq:zeta-poles}
\end{equation}
for positive constants $C_{\mathrm{DI}}$ and $C_{\mathrm{SPADE}}$. The rightmost pole is $z=-1/2$ in both cases, but the multiplicity differs:
\begin{equation}
(\lambda_{\mathrm{DI}},m_{\mathrm{DI}})=\left(\frac12,2\right),
\qquad
(\lambda_{\mathrm{SPADE}},m_{\mathrm{SPADE}})=\left(\frac12,1\right).
\label{eq:RLCT-aligned}
\end{equation}

This result concerns a different asymptotic object from the fixed-parameter Stein exponents of Sec.~II. The prior is fixed as $n\to\infty$; it is the marginal-likelihood integral that becomes increasingly localized near the KL-zero set. Consequently, alternatives that are arbitrarily close to the one-source boundary dominate the prior-predictive asymptotics even though any fixed separated alternative has an exponential error exponent.

Substituting Eq.~\eqref{eq:RLCT-aligned} into Eq.~\eqref{eq:singular-Fn}, we obtain
\begin{align}
F_n^{\mathrm{DI}}
&=
\frac{1}{2}\log n-\log\log n+O_p(1),\nonumber\\
F_n^{\mathrm{SPADE}}
&=
\frac{1}{2}\log n+O(1).
\label{eq:aligned-free-energies}
\end{align}
The stronger deterministic remainder for aligned SPADE follows because under $H_0$ every detected photon occupies $q=0$ almost surely, so its Bayes free energy has no sample-to-sample fluctuation.

Applying the change-of-measure relation from Sec.~III to the prior-mixture alternatives gives
\begin{align}
-\log \bar\beta_{n,\phi}^{\mathrm{DI},*}(\alpha)
&=
\frac{1}{2}\log n-\log\log n+O(1),
\label{eq:aligned-beta-DI}
\\
-\log \bar\beta_{n,\phi}^{\mathrm{SPADE},*}(\alpha)
&=
\frac{1}{2}\log n+O(1).
\label{eq:aligned-beta-SPADE}
\end{align}
Thus, within the smooth-positive prior class in the physical $(\epsilon,s)$ coordinates, aligned SPADE has a subleading logarithmic distinction in prior-mixture type-II decay. This does not contradict the much larger fixed-parameter Stein advantage: the two quantities answer different questions, and the prior mixture is dominated by the nearly indistinguishable part of the alternative model. Boundary-weighted priors can alter the pole data, as quantified in Appendix~E.

As a numerical sanity check, Appendix~A compares the exact finite-$n$ Bayes free energy for aligned direct imaging with the corresponding centered local likelihood approximation under the same Monte Carlo samples and several bounded prior windows. This check validates the local likelihood reduction, rather than serving as a numerical estimate of the RLCT itself.

\subsection{Misaligned case: local singular geometry and a fixed-offset finite-\texorpdfstring{$n$}{n} stress test}

To expose the misaligned singular structure in the simplest physically meaningful form, we study the binary-SPADE statistic that records whether a photon remains in the lowest-order mode. Appendix~C shows that near ideal alignment this binary statistic retains the leading leakage contrast of Hermite--Gaussian mode counting.

Let
\[
Y_i=
\begin{cases}
1, & \text{if the $i$th photon is detected in the mode $q=0$},\\
0, & \text{otherwise}.
\end{cases}
\]
Under a sorter offset $\theta$,
\begin{equation}
H_0:\quad Y_i\sim\mathrm{Bernoulli}\!\bigl(p_0(\theta)\bigr),
\end{equation}
while under the fixed-primary two-source alternative,
\begin{equation}
H_1:\quad Y_i\sim\mathrm{Bernoulli}\!\bigl(q_{\epsilon,s,\theta}\bigr),
\qquad
q_{\epsilon,s,\theta}=(1-\epsilon)p_0(\theta)+\epsilon p_s(\theta),
\end{equation}
with
\[
p_0(\theta)=e^{-\gamma^2},\qquad
p_s(\theta)=e^{-(\gamma-g_s)^2},\qquad
\gamma=\frac{\theta}{2\sigma},\quad g_s=\frac{s}{2\sigma}.
\]
Writing $\Delta=q_{\epsilon,s,\theta}-p_0(\theta)$ gives the exact form
\begin{equation}
\Delta
=
\epsilon p_0(\theta)\left(e^{-\eta}-1\right),
\qquad
\eta:=\frac{s(s-2\theta)}{4\sigma^2}.
\label{eq:misaligned-fixed-primary-delta}
\end{equation}
The global KL-zero set of this binary model is therefore
\begin{equation}
W_0=\{\epsilon=0\}\cup\{s=0\}\cup\{s=2\theta\}.
\label{eq:misaligned-zero-set}
\end{equation}
For a fixed nonzero $\theta$, the additional component $s=2\theta$ lies outside a sufficiently small neighborhood of $(\epsilon,s)=(0,0)$. Hence, within the neighborhood used for the local zeta analysis, the KL-zero set is still $\{\epsilon=0\}\cup\{s=0\}$. Appendix~C shows that the branch $s=2\theta$ is not created by binary coarse graining: the full Hermite--Gaussian mode-counting distribution has the same reflection symmetry about the sorter axis.

Expanding Eq.~\eqref{eq:misaligned-fixed-primary-delta} around $s=0$ gives
\[
\Delta
=
\epsilon p_0(\theta)
\left[2\gamma g_s+(2\gamma^2-1)g_s^2+O(g_s^3)\right].
\]
For fixed nonzero $\theta$, the first term is linear in $s$, and the Bernoulli KL divergence becomes
\begin{equation}
D^{\mathrm{mis,fp}}_{\mathrm{bSPADE}}(\epsilon,s,\theta)
=
\frac{\epsilon^2p_0(\theta)\theta^2}{8\sigma^4[1-p_0(\theta)]}s^2+O(s^3).
\label{eq:misaligned-fixed-primary-bspade-KL}
\end{equation}
Equation~\eqref{eq:misaligned-fixed-primary-bspade-KL} is a strict local expansion as $s\to0$ at fixed $\epsilon$ and fixed $\theta\ne0$. For the joint local geometry relevant to the zeta function, Eq.~\eqref{eq:misaligned-fixed-primary-delta} gives the stronger factorization
\begin{equation}
D^{\mathrm{mis,fp}}_{\mathrm{bSPADE}}(\epsilon,s,\theta)
=\epsilon^2 s^2 h_\theta(\epsilon,s),
\qquad
h_\theta(0,0)
=\frac{p_0(\theta)\theta^2}{8\sigma^4[1-p_0(\theta)]}>0,
\label{eq:misaligned-local-factorization}
\end{equation}
where $h_\theta$ is analytic and positive in a sufficiently small neighborhood of $(0,0)$ for fixed $\theta\ne0$. Thus the local KL function is genuinely of normal-crossing order $\epsilon^2s^2$ when $\epsilon$ and $s$ vary jointly. The explicit quadratic approximation in Eq.~\eqref{eq:misaligned-fixed-primary-bspade-KL} additionally requires both the linear term in $\Delta$ to dominate and $|\Delta|\ll1-p_0(\theta)$; for $|\theta|\ll\sigma$ these conditions reduce approximately to
\begin{equation}
|s|\ll \min\!\left(2|\theta|,\frac{|\theta|}{2\epsilon}\right).
\label{eq:misaligned-validity}
\end{equation}
The pointwise finite-$n$ calculations below use the exact Bernoulli law and do not rely on this local approximation over the plotted range.

In the same fixed-primary convention, direct imaging satisfies
\begin{equation}
D^{\mathrm{mis,fp}}_{\mathrm{DI}}(\epsilon,s)
=
\frac{\epsilon^2s^2}{2\sigma^2}+O(\epsilon^2s^4).
\label{eq:misaligned-fixed-primary-DI-KL}
\end{equation}
Together with Eq.~\eqref{eq:misaligned-fixed-primary-DI-KL}, the factorization in Eq.~\eqref{eq:misaligned-local-factorization} shows that both local KL functions have the same normal-crossing order $\epsilon^2s^2$. For a prior smooth and strictly positive in the physical $(\epsilon,s)$ coordinates, the fixed-offset binary-SPADE model therefore has
\begin{equation}
(\lambda,m)^{\mathrm{mis,fp}}_{\mathrm{bSPADE}}
=\left(\frac12,2\right)
=(\lambda,m)_{\mathrm{DI}}.
\label{eq:misaligned-RLCT}
\end{equation}
This connects the aligned and misaligned calculations directly: the ideal aligned support mismatch produces the $\epsilon s^2$ monomial and multiplicity $m=1$, whereas any fixed nonzero offset gives common support under the null and restores a quadratic perturbation, $\epsilon^2s^2$, for the binary statistic.

\begin{figure*}[t]
  \centering
  \includegraphics[width=\textwidth]{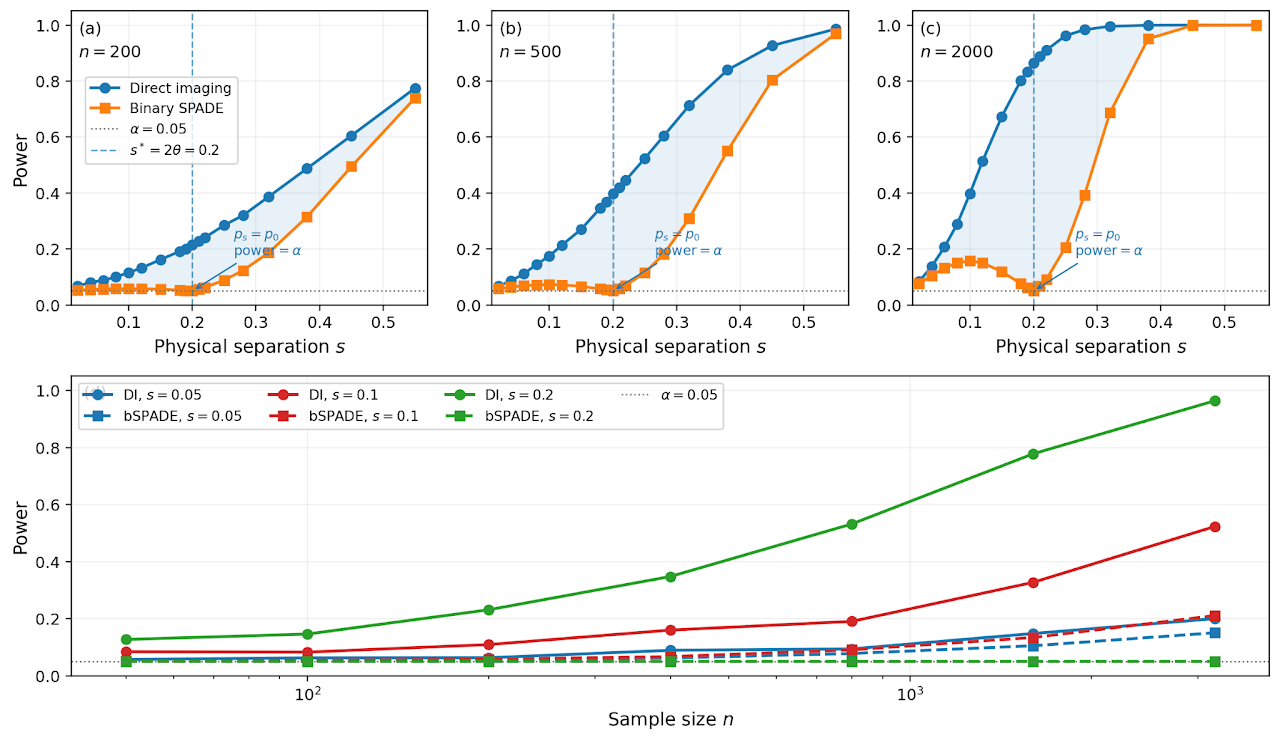}
  \caption{Pointwise finite-$n$ power for ideal direct imaging and the fixed-offset binary-SPADE statistic at common specified parameter values. Panels~(a)--(c): power versus physical separation $s$ for $n=200,500,2000$; panel (d): power versus sample size $n$ for fixed separations $s=0.05,0.10,0.20$. Parameters are $\sigma=1$, $\epsilon=0.3$, $\theta=0.1$, and $\alpha=0.05$. At each displayed $s$, the simple-versus-simple Neyman--Pearson test is calibrated for that alternative. Binary-SPADE power is computed exactly from the binomial model using boundary randomization to obtain exact size $\alpha$; direct-imaging power is evaluated by common-random-number Monte Carlo. The dashed vertical line marks $s^*=2\theta=0.2$, where the null and alternative binary-SPADE laws coincide, so the exact size-$\alpha$ test has power $\alpha$. The blind point reflects the underlying reflection symmetry of fixed-offset Hermite--Gaussian mode counting. On the plotted grids direct imaging lies above this fixed-offset binary statistic; no adaptive recentering stage is included.}
  \label{fig:misaligned_power_physical}
\end{figure*}

The leading binary-SPADE coefficient relative to direct imaging is
\begin{equation}
R_{\mathrm{fp}}(\theta)
:=
\frac{C_{\mathrm{bSPADE}}(\theta)}{C_{\mathrm{DI}}}
=
\frac{\tau_0e^{-\tau_0}}{1-e^{-\tau_0}},
\qquad
\tau_0:=\frac{\theta^2}{4\sigma^2}.
\label{eq:misaligned-coefficient-ratio}
\end{equation}
For the value $\theta=0.1\sigma$ used in Fig.~\ref{fig:misaligned_power_physical}, $R_{\mathrm{fp}}\simeq0.99875$. Hence the pronounced finite-$n$ gap in the figure is not a consequence of a large difference in the leading $s\to0$ coefficient; it instead reflects higher-order, nonlocal structure beyond the strict local expansion, most visibly the reflection-induced zero branch at $s=2\theta$.

For fixed $\epsilon>0$ and fixed $\theta\ne0$, $nD=O(1)$ corresponds to
\begin{equation}
s=O(n^{-1/2})
\qquad\text{(fixed-primary convention),}
\label{eq:misaligned-fixed-primary-scale}
\end{equation}
while the more intrinsic local condition is $\epsilon s=O(n^{-1/2})$ when both parameters are allowed to vary.

A useful heuristic measure of when the aligned support structure ceases to describe the observed null counts is the expected number of null photons outside the lowest mode,
\[
n[1-p_0(\theta)]\simeq \frac{n\theta^2}{4\sigma^2}
\qquad (|\theta|\ll\sigma).
\]
The crossover $n[1-p_0]\sim1$ therefore occurs around
\begin{equation}
n_c\sim\frac{4\sigma^2}{\theta^2}.
\label{eq:misaligned-crossover}
\end{equation}
For $\theta=0.1\sigma$, this gives $n_c\sim400$, between the smallest and larger sample sizes shown in Fig.~\ref{fig:misaligned_power_physical}. This is a heuristic crossover scale, not a sharp phase boundary.

For comparison, under the centroid-preserving convention $x_1=-\epsilon s$ and $x_2=(1-\epsilon)s$, the first-order translation signal cancels. For generic fixed offsets satisfying $2\gamma^2\ne1$, Appendix~B gives
\begin{align}
D^{\mathrm{mis,cf}}_{\mathrm{bSPADE}}(\epsilon,s,\theta)
&\asymp \epsilon^2(1-\epsilon)^2s^4,\nonumber\\
D^{\mathrm{mis,cf}}_{\mathrm{DI}}(\epsilon,s)
&\asymp \epsilon^2(1-\epsilon)^2s^4,
\label{eq:misaligned-centroid-fixed-KL}
\end{align}
and, for fixed $\epsilon$,
\begin{equation}
s=O(n^{-1/4})
\qquad\text{(centroid-preserving convention).}
\label{eq:misaligned-centroid-fixed-scale}
\end{equation}
The local order is therefore a property of the nuisance direction rather than of the receiver label alone; Appendix~B records the corresponding leading coefficients.

\paragraph*{Relation to adaptive alignment.}
Adaptive and hybrid schemes that estimate the centroid and recenter the mode sorter have been studied in Refs.~\cite{Tsang2016,Grace2020,OzerGrace2024}. We do not model that alignment stage here; $\theta$ is held fixed to isolate the singular structure associated with a residual offset, while direct imaging is kept idealized.

To make the fixed-offset comparison explicit, we computed pointwise finite-$n$ Neyman--Pearson powers at the same specified values of $n,\epsilon,s,\theta$, and $\alpha$ for the two observation models. For each value of $s$, the simple-versus-simple test is optimized separately. Binary-SPADE power is computed exactly from the binomial model with boundary randomization to obtain exact size $\alpha$; direct-imaging power is evaluated by common-random-number Monte Carlo. Figure~\ref{fig:misaligned_power_physical} uses the representative dimensionless stress-test value $\theta=0.1\sigma$ rather than a calibration to a particular experiment.

On the plotted grids, ideal direct imaging lies above the fixed-offset binary-SPADE statistic. The exact form in Eq.~\eqref{eq:misaligned-fixed-primary-delta} also explains the characteristic nonmonotonic binary curve: $\eta(s)=\eta(2\theta-s)$, so the Bernoulli alternative and hence its pointwise Neyman--Pearson power are reflection-symmetric about $s=\theta$. At
\begin{equation}
s^*=2\theta,
\label{eq:bspade_blindspot}
\end{equation}
one has $\Delta=0$, and every binary-SPADE test has power equal to its own size. More generally, Appendix~C shows that at this point the entire full Hermite--Gaussian mode-counting law of the secondary source equals that of the null source; the blind branch is therefore a property of fixed-offset HG mode counting, not an artifact of the binary reduction.

\section{Conclusion}

The question motivating this work is not whether ideal SPADE can outperform direct imaging for a specified two-source alternative---that fixed-parameter advantage is already captured by the usual Stein exponents---but how that conclusion changes when the alternative approaches a non-identifiable one-source boundary and its parameters are not known a priori. In that setting the prior-predictive problem is controlled by a neighborhood of the Kullback--Leibler zero set rather than by any single separated alternative. Singular learning theory makes this distinction explicit: for priors smooth and strictly positive in the physical $(\epsilon,s)$ coordinates, aligned direct imaging and SPADE have the same RLCT $\lambda=1/2$ but different multiplicities, yielding different subleading model-evidence asymptotics even though their fixed-parameter Stein exponents differ already at leading order.

The misalignment analysis clarifies where this singular distinction comes from. Ideal alignment gives SPADE an exact support mismatch under the null and the local monomial $K\asymp\epsilon s^2$. Any fixed nonzero sorter offset removes that support mismatch; in the fixed-primary binary-SPADE model the local behavior becomes $K\asymp\epsilon^2s^2$, so under the same smooth-positive prior class the pole data become $(\lambda,m)=(1/2,2)$, as for direct imaging. At the same time, the full Hermite--Gaussian mode-counting law has the reflection symmetry $x\mapsto2\theta-x$, which produces the additional KL-zero branch $s=2\theta$ and the exact finite-$n$ power collapse exposed by the binary stress test. The centroid-preserving calculation further shows that the local separation order depends on the nuisance direction being probed, not on the receiver label alone.

Taken together, these results suggest a more precise way to interpret finite-photon optimality near a singular boundary. The relevant asymptotics are determined jointly by the local prior measure, the support geometry induced by the measurement, and the nuisance direction along which the alternative approaches the KL-zero set. A receiver can therefore be optimal in the fixed-parameter Stein sense while exhibiting qualitatively different prior-predictive or finite-$n$ behavior once alignment and source parameters are not fixed. This is the main conceptual link between the singular-learning analysis and the robustness question raised in the Introduction.

The next step is to test how far this geometric picture survives in more realistic receiver models. Mode-sorting crosstalk populates modes that are forbidden under ideal SPADE and is known to degrade sub-Rayleigh performance~\cite{Gessner2020,Linowski2023}; because the resulting local KL monomial depends on the detailed error channel, it need not coincide with the misalignment case studied here. Adaptive two-stage receivers pose a complementary problem: part of a finite photon budget is first spent estimating and recentering the source before mode sorting. An end-to-end analysis that combines this resource allocation with both prior-predictive evidence and finite-$n$ Neyman--Pearson power would provide the natural next comparison between idealized singular advantages and experimentally robust discrimination.

\begin{acknowledgments}
The author is grateful to Mankei Tsang for hosting the 3rd Workshop on Quantum Metrology and Quantum-Inspired Superresolution (QISR26) and for insightful comments and discussions that helped clarify the role of prior dependence and the scope of the misalignment analysis. The author also thanks the participants of QISR26 for helpful discussions.
\end{acknowledgments}

\section*{Author contribution statement}

The author conceived the study, developed the analytical framework, carried out the derivations and numerical analyses, and wrote the manuscript. During the preparation of version 3, OpenAI's ChatGPT was used as an assistive tool for algebraic-consistency checks, English-language polishing of revised passages, and generation of auxiliary code used for numerical and algebraic verification. All scientific judgments and final responsibility for the derivations, calculations, citations, and text remain with the author.

\appendix

\section{Numerical validation of the aligned direct-imaging local approximation across multiple bounded prior windows}
To assess how well the small-$(\epsilon,s)$ likelihood reduction describes the finite-$n$ regime, we compare under $H_0$ the exact finite-$n$ Bayes free energy for aligned direct imaging with the corresponding local likelihood approximation. For visual comparison with the analytic singular expansion, we center both quantities by
\[
\frac{1}{2}\log n-\log\log n.
\]

More precisely, for each Monte Carlo realization we evaluate
\begin{eqnarray*}
\widetilde F_n^{\mathrm{exact}}
&:=&
F_n^{\mathrm{exact}}
-
\left(
\frac{1}{2}\log n-\log\log n
\right),
\\
\widetilde F_n^{\mathrm{local}}
&:=&
F_n^{\mathrm{local}}
-
\left(
\frac{1}{2}\log n-\log\log n
\right),
\end{eqnarray*}

and compare their empirical distributions.

The purpose of this numerical check is to verify that the local likelihood model reproduces the exact finite-$n$ Bayes free energy, in the central part of its distribution, across several bounded prior windows within the weak-companion, sub-Rayleigh regime. It is a check of the local reduction used in the analytic calculation, not a direct numerical estimator of the RLCT or multiplicity. Using several windows also tests that the agreement is not an artifact of a single ad hoc parameter choice.
Accordingly, instead of fixing only one prior window, we consider three bounded uniform priors on $(\epsilon,s)$:
\[
(\epsilon_{\max},s_{\max}/\sigma)
=
(0.10,0.25),\quad
(0.10,0.30),\quad
(0.15,0.40).
\]

All three windows remain in the local regime in the sense that the leading-order direct-imaging Kullback--Leibler scale
\[
D_{\max}^{\mathrm{lead}}
:=
\frac{\epsilon_{\max}^2 s_{\max}^2}{2\sigma^2}
\]
is small throughout; the corresponding values are shown in Fig.~2. For each window and each sample size
\[
n=32,64,128,256,512,1024,2048,
\]

we generated Monte Carlo samples under $H_0$ with $X_i\sim \mathcal N(0,\sigma^2)$ using a fixed random seed ($12345$).

To make the local approximation explicit, write the exact direct-imaging log likelihood ratio as
\[
\ell_n(\epsilon,s)
=
\sum_{i=1}^n
\log\!\left[
1+\epsilon\left(
\exp\!\left(\frac{X_i s}{\sigma^2}-\frac{s^2}{2\sigma^2}\right)-1
\right)
\right].
\]
Under $H_0$, define
\[
Z_n:=\frac{1}{\sigma\sqrt n}\sum_{i=1}^n X_i,
\qquad Z_n\Rightarrow N(0,1),
\]
and $u:=\sqrt n\,\epsilon s/\sigma$. In the joint local region with $\epsilon\to0$ and $s\to0$ (hence approaching $W_0^{\mathrm{al}}$) and with $u=O(1)$, the leading local likelihood ratio is
\begin{equation}
\ell_n^{\mathrm{local}}(\epsilon,s)
=
u Z_n-\frac12u^2+o_p(1),
\qquad u=\frac{\sqrt n\,\epsilon s}{\sigma}.
\label{eq:A-local-LR}
\end{equation}
For a uniform prior on $0<\epsilon<A$ and $0<s<B$, this gives
\begin{align}
\Lambda_n^{\mathrm{local}}
&=
\frac{1}{AB}
\int_0^A\!d\epsilon\int_0^B\!ds\,
\exp\!\left(uZ_n-\frac12u^2\right) \\
&=
\frac{1}{AB}
\int_0^{AB}\!dr\,
\log\!\frac{AB}{r}
\exp\!\left(
\frac{\sqrt n\,r}{\sigma}Z_n
-\frac{nr^2}{2\sigma^2}
\right),
\label{eq:A-local-integral}
\end{align}
where $r=\epsilon s$. This is the one-dimensional local integral used below.

For each realization, the exact finite-$n$ Bayes free energy was evaluated by numerical quadrature of the bounded prior integral, while the local approximation was evaluated from Eq.~\eqref{eq:A-local-integral}. In the computations shown in Fig.~2 we set $\sigma=1$ and used Gauss--Legendre quadrature for the exact bounded-prior integral. The comparison is therefore performed under the same bounded prior window and the same Monte Carlo sample for both the exact and local quantities.

Figure~2 shows, for each prior window, the empirical median and central $10\%$--$90\%$ quantile band of the centered exact finite-$n$ Bayes free energy together with those of the centered local model. Across all three prior windows, the local model tracks the central part of the exact finite-$n$ distribution well. As expected, the agreement is strongest for the smaller prior windows and becomes slightly less sharp as the window is enlarged. This numerical check validates the small-$(\epsilon,s)$ likelihood approximation used to obtain the local monomial $K\asymp\epsilon^2s^2$. It should not, by itself, be read as a numerical estimate of the RLCT or as evidence that the centered quantity has already reached its limiting $O_p(1)$ distribution at the displayed sample sizes; the latter is an asymptotic statement derived analytically from the zeta-function pole structure.

\begin{figure*}[t]
 \centering
 \includegraphics[width=\textwidth]{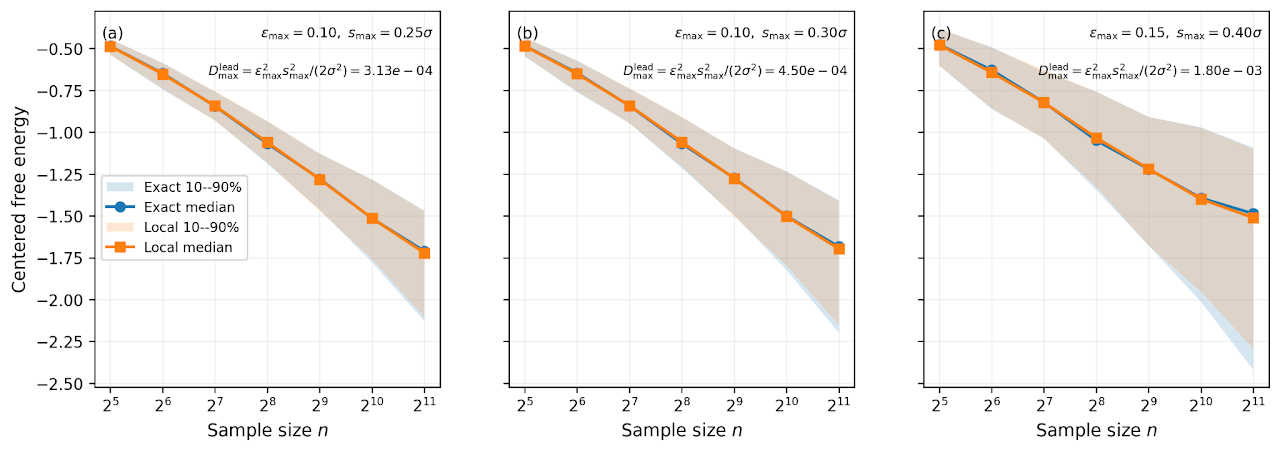}
 \caption{
 Numerical validation of the aligned direct-imaging local approximation across multiple bounded prior windows. For each panel, Monte Carlo samples are generated under $H_0$ with $X_i\sim \mathcal N(0,\sigma^2)$, and the centered exact finite-$n$ Bayes free energy $\widetilde F_n^{\mathrm{exact}} = F_n^{\mathrm{exact}} - \left( \frac{1}{2}\log n-\log\log n \right)$ is compared with the corresponding centered local approximation $\widetilde F_n^{\mathrm{local}} = F_n^{\mathrm{local}} - \left( \frac{1}{2}\log n-\log\log n \right)$.
 Solid lines show the empirical medians and shaded bands show the empirical $10\%$--$90\%$ quantile ranges. The three panels correspond to the bounded uniform prior windows $(\epsilon_{\max},s_{\max}/\sigma)=(0.10,0.25)$, $(0.10,0.30)$, and $(0.15,0.40)$. In each panel, $D_{\max}^{\mathrm{lead}} = \frac{\epsilon_{\max}^2 s_{\max}^2}{2\sigma^2}$ denotes the largest leading-order direct-imaging Kullback--Leibler scale inside the corresponding prior window. The agreement across several prior windows shows that the local likelihood approximation reproduces the central part of the exact finite-$n$ distribution over the displayed parameter windows; the figure is not used to estimate the RLCT or the limiting centered distribution.
 }
 \label{fig:appendixA_multwindow_validation}
\end{figure*}

\section{Local statistics under fixed-primary and centroid-preserving parameterizations}

This appendix supplies the local expansions used in Sec.~IV~B while keeping the source-position convention explicit. We use the notation introduced there and keep $\theta\ne0$ fixed, so $0<p_0(\theta)<1$; the aligned support structure is treated separately in Sec.~IV~A.

For a Bernoulli null parameter $p_0$ and alternative parameter $p_0+\Delta$, the single-observation log-likelihood ratio is
\begin{equation}
\ell_i
=
X_i\log\frac{p_0+\Delta}{p_0}
+(1-X_i)\log\frac{1-p_0-\Delta}{1-p_0}.
\label{eq:B-generic-ell}
\end{equation}
Its second-order expansion is
\begin{align}
\ell_i
&=
\frac{X_i-p_0}{p_0(1-p_0)}\Delta \nonumber\\
&\quad-
\frac12\left[
\frac{X_i}{p_0^2}
+
\frac{1-X_i}{(1-p_0)^2}
\right]\Delta^2
+O(\Delta^3).
\label{eq:B-generic-pointwise}
\end{align}
Defining
\begin{equation}
\xi_n:=\frac{\sqrt n(\bar X_n-p_0)}{\sqrt{p_0(1-p_0)}},
\label{eq:B-xi}
\end{equation}
one has $\xi_n\Rightarrow N(0,1)$ under $H_0$, and whenever $\sqrt n\Delta=O(1)$,
\begin{equation}
\sum_{i=1}^n\ell_i
=
t_n\xi_n-\frac12t_n^2+o_p(1),
\qquad
t_n:=\frac{\sqrt n\Delta}{\sqrt{p_0(1-p_0)}}.
\label{eq:B-generic-LAN}
\end{equation}
For later use,
\begin{equation}
D\!\left(\operatorname{Ber}(p_0)\middle\|\operatorname{Ber}(p_0+\Delta)\right)
=
\frac{\Delta^2}{2p_0(1-p_0)}+O(\Delta^3).
\label{eq:B-Bernoulli-KL}
\end{equation}

\subsection*{Fixed-primary convention}
For the model defined in Sec.~IV~B, the exact perturbation is already given in Eq.~\eqref{eq:misaligned-fixed-primary-delta}. Expanding it at $s=0$, with $\gamma=\theta/(2\sigma)$ and $\beta=s/(2\sigma)$, gives
\begin{equation}
\Delta^{\mathrm{fp}}
=
\epsilon p_0(\theta)
\left[2\gamma\beta+(2\gamma^2-1)\beta^2+O(\beta^3)\right].
\label{eq:B-fp-delta}
\end{equation}
Using Eq.~\eqref{eq:B-Bernoulli-KL},
\begin{equation}
D_{\mathrm{bSPADE}}^{\mathrm{fp}}(\epsilon,s,\theta)
=
\frac{\epsilon^2p_0(\theta)\theta^2}{8\sigma^4[1-p_0(\theta)]}s^2+O(s^3).
\label{eq:B-fp-bspade-KL}
\end{equation}
For direct imaging,
\begin{equation}
D_{\mathrm{DI}}^{\mathrm{fp}}(\epsilon,s)
=
\frac{\epsilon^2s^2}{2\sigma^2}+O(\epsilon^2s^4).
\label{eq:B-fp-DI-KL}
\end{equation}
The ratio of the leading coefficients is
\begin{equation}
\frac{C_{\mathrm{bSPADE}}(\theta)}{C_{\mathrm{DI}}}
=
\frac{\tau_0e^{-\tau_0}}{1-e^{-\tau_0}},
\qquad \tau_0=\frac{\theta^2}{4\sigma^2},
\label{eq:B-fp-ratio}
\end{equation}
which approaches unity as $\theta\to0$ through nonzero values. The quantitative validity conditions for the local expansion are stated in Sec.~IV~B.

\subsection*{Centroid-preserving convention}
Under the centroid-preserving convention,
\begin{equation}
x_1=-\epsilon s,
\qquad
x_2=(1-\epsilon)s,
\label{eq:B-cf-positions}
\end{equation}
so that $(1-\epsilon)x_1+\epsilon x_2=0$. The alternative Bernoulli parameter is
\begin{equation}
q_{\epsilon,s,\theta}^{\mathrm{cf}}
=(1-\epsilon)e^{-(\gamma+\epsilon\beta)^2}
+\epsilon e^{-[\gamma-(1-\epsilon)\beta]^2}.
\label{eq:B-cf-q}
\end{equation}
The linear displacement terms cancel, and
\begin{equation}
\Delta^{\mathrm{cf}}
=
p_0(\theta)(2\gamma^2-1)\epsilon(1-\epsilon)\beta^2+O(\beta^3).
\label{eq:B-cf-delta}
\end{equation}
For a generic fixed offset with $2\gamma^2\ne1$,
\begin{equation}
D_{\mathrm{bSPADE}}^{\mathrm{cf}}(\epsilon,s,\theta)
=
\frac{p_0(\theta)(2\gamma^2-1)^2}{32\sigma^4[1-p_0(\theta)]}
\epsilon^2(1-\epsilon)^2s^4+O(s^5).
\label{eq:B-cf-bspade-KL}
\end{equation}
For direct imaging,
\begin{equation}
p_1(x)-\phi_\sigma(x)
=
\frac{\epsilon(1-\epsilon)s^2}{2}\phi_\sigma''(x)+O(s^3),
\label{eq:B-cf-DI-density-expansion}
\end{equation}
and since
\[
\int\frac{[\phi_\sigma''(x)]^2}{\phi_\sigma(x)}\,dx=\frac{2}{\sigma^4},
\]
one obtains
\begin{equation}
D_{\mathrm{DI}}^{\mathrm{cf}}(\epsilon,s)
=
\frac{\epsilon^2(1-\epsilon)^2s^4}{4\sigma^4}+O(s^5).
\label{eq:B-cf-DI-KL}
\end{equation}
The ratio of the leading coefficients is therefore
\begin{equation}
\frac{C_{\mathrm{bSPADE}}^{\mathrm{cf}}}{C_{\mathrm{DI}}^{\mathrm{cf}}}
=
\frac{p_0(\theta)(2\gamma^2-1)^2}{8[1-p_0(\theta)]}.
\label{eq:B-cf-ratio}
\end{equation}
For the numerical offset $\theta=0.1\sigma$ used in Fig.~\ref{fig:misaligned_power_physical}, this ratio is approximately $49.4$. This coefficient comparison belongs to the centroid-preserving local experiment and is not the convention used for Fig.~\ref{fig:misaligned_power_physical}. At the exceptional offset $\theta=\sqrt2\,\sigma$ the displayed $s^4$ coefficient vanishes, so a higher-order expansion is required.

Equations~\eqref{eq:B-fp-bspade-KL}--\eqref{eq:B-cf-DI-KL} show that the $n^{-1/2}$ and $n^{-1/4}$ separation scales arise from different source-position conventions. They therefore cannot be used as a receiver ranking by combining results from different nuisance directions.

\section{Full mode-counting symmetry and the leading binary contrast}
This appendix records two properties of Gaussian Hermite--Gaussian (HG) mode counting used in Sec.~IV~B. The coordinates $x$, $\theta$, and the fixed-primary displacement $s$ are signed transverse coordinates; the physical source separation is $|s|$. If the HG sorter is centered at $\theta$ and a point source is located at $x$, the mode occupation probabilities are
\begin{equation}
P_\theta(q|x)
=
e^{-\tau_x}\frac{\tau_x^q}{q!},
\qquad
\tau_x:=\frac{(x-\theta)^2}{4\sigma^2},
\qquad q=0,1,2,\ldots .
\label{eq:C1-poisson}
\end{equation}
Thus the full mode-counting law depends on the source position only through $|x-\theta|$. It follows immediately that
\begin{equation}
P_\theta(q|x)=P_\theta(q|2\theta-x)
\qquad\text{for every }q.
\label{eq:C-reflection}
\end{equation}
In the fixed-primary model, the null source is at $x=0$. Hence a secondary source at $s=2\theta$ has exactly the same full HG mode-counting distribution as the null source. The additional KL-zero branch identified in Sec.~IV~B is therefore a reflection property of the full fixed-offset HG measurement, not an artifact of binary coarse graining.

For the leading local contrast, set $\theta=0$ and $x=s$. Then $\tau=s^2/(4\sigma^2)$ and
\begin{equation}
P_0(0|s)=e^{-\tau}=1-\tau+O(\tau^2),
\label{eq:C1-q0}
\end{equation}
so the total leakage probability out of the lowest-order mode is
\begin{equation}
1-P_0(0|s)=\tau+O(\tau^2)=\frac{s^2}{4\sigma^2}+O(s^4).
\label{eq:C1-leakage}
\end{equation}
The first excited mode carries
\begin{equation}
P_0(1|s)=e^{-\tau}\tau=\tau+O(\tau^2),
\label{eq:C1-q1}
\end{equation}
whereas
\begin{equation}
\sum_{q\ge2}P_0(q|s)=O(\tau^2)=O(s^4).
\label{eq:C1-higher}
\end{equation}
Thus, near ideal alignment, the binary statistic $q=0$ versus $q\ge1$ retains the complete leading $O(s^2)$ leakage contrast, while the redistribution among higher modes enters at $O(s^4)$. This is the sense in which binary-SPADE provides a minimal local model for the misalignment calculation in the main text.

\section{A minimal review of singular learning theory for the present problem}
For the convenience of readers unfamiliar with singular learning theory, we briefly summarize here the part of the theory that is used in the main text. We restrict attention to the quantities that enter the present one-versus-two-source discrimination problem and omit topics, such as generalization error and singular fluctuation, that are not needed below. General references include Watanabe's monograph~\cite{Watanabe2009} and the recent review~\cite{Watanabe2024Review}; see also Refs.~\cite{KariyaWatanabe2022,KariyaWatanabe2020} for singular hypothesis testing in related settings.
\paragraph*{Singular versus regular models.}

In a regular parametric model, the map from the parameter $w$ to the observable probability distribution $p(\,\cdot\,|w)$ is locally one-to-one, and the Fisher information matrix is positive definite at the parameter representing the data-generating distribution. In that case, the posterior distribution is asymptotically Gaussian, and the familiar regular asymptotic theory applies. By contrast, if different parameter values yield the same observable distribution, the Fisher information degenerates and the posterior need not be Gaussian. Such a model is called \emph{singular}. Mixture models and latent-variable models are typical examples~\cite{Hartigan1985,ChenLi2009,Watanabe2009}. The one-versus-two-source discrimination problem considered in this paper becomes singular because the one-source distribution is represented non-identifiably along a boundary subset of the two-source family.

\paragraph*{Bayes factor, marginal likelihood, and free energy.}
Let $Y^n=(Y_1,\dots,Y_n)$ be independent observations, let $p_0(y)$ denote the null distribution, and let $p(y|w)$ denote the alternative model with prior density $\varphi(w)$ on a parameter space $W$. The marginal likelihood under the alternative is
\[
m_1(Y^n)=\int_W \prod_{i=1}^n p(Y_i|w)\,\varphi(w)\,dw,
\]

whereas under the null
\[
m_0(Y^n)=\prod_{i=1}^n p_0(Y_i).
\]

The Bayes factor is
\[
\Lambda_n(Y^n)=\frac{m_1(Y^n)}{m_0(Y^n)},
\]

and the minus log Bayes factor
\[
F_n(Y^n):=-\log \Lambda_n(Y^n)
\]

is the free-energy difference relevant for Bayesian hypothesis testing. In the present paper, this is the natural test statistic because, once the composite alternative is equipped with a prior, the null-versus-alternative problem becomes an ordinary simple-versus-simple comparison between $m_0$ and $m_1$; see Sec.~III.

\paragraph*{The Kullback--Leibler function and the zeta function.}

The asymptotic behavior of the marginal likelihood is governed by the Kullback--Leibler function
\[
K(w):=\int p_0(y)\log\frac{p_0(y)}{p(y|w)}\,d\nu(y),
\]

where $\nu$ is the underlying measure on the observation space. In regular problems, $K(w)$ has an isolated nondegenerate minimum and Laplace's method applies. In singular problems, however, the zero set
\[
W_0:=\{w\in W\,;\,K(w)=0\}
\]

typically contains singularities or higher-dimensional subsets.

A central object of singular learning theory is the zeta function
\[
\zeta(z):=\int_W K(w)^z \varphi(w)\,dw.
\]

Under the standard analyticity assumptions used in singular learning theory, $\zeta(z)$ extends meromorphically, and its poles determine the asymptotics of the marginal likelihood~\cite{Watanabe2009,Watanabe2024Review}.

\paragraph*{Real log canonical threshold and multiplicity.}

Let $-\lambda$ be the largest pole of $\zeta(z)$ and let $m$ be its order. Then $\lambda$ is called the \emph{real log canonical threshold} (RLCT), and $m$ its \emph{multiplicity}. In regular models of parameter dimension $d$, one has
\[
\lambda=\frac{d}{2},\qquad m=1.
\]

In singular models, however, $\lambda$ and $m$ are birational invariants of the pair $(K,\varphi)$ and generally differ from the regular values~\cite{Watanabe2009}. Intuitively, $\lambda$ plays the role of an effective dimension for the leading free-energy growth, while $m$ controls the subleading $\log\log n$ correction.

\paragraph*{The asymptotic form of the free energy.}

A basic result of singular learning theory states that, in the realizable case and under suitable analyticity assumptions,

\begin{eqnarray*}
-\log m_1(Y^n)
&=&
-\sum_{i=1}^n \log p_0(Y_i)
+
\lambda \log n \\
&-&
(m-1)\log\log n
+
O_p(1).
\end{eqnarray*}

Hence the free-energy difference satisfies

\[
F_n(Y^n)
=
\lambda \log n
-
(m-1)\log\log n
+
O_p(1).
\]

This is the formula used in Eq.~\eqref{eq:singular-Fn} of the main text. It shows that, in singular problems, the leading deterministic behavior is logarithmic in $n$, rather than linear in $n$, when the null distribution lies on the boundary of the alternative model.

\paragraph*{Why the zeta function appears.}

Very briefly, the reason is as follows. A key point of resolution of singularities is that, after a suitable local change of variables, the singular structure of the Kullback--Leibler function can be reduced to a normal-crossing form. More precisely, the content relevant here is that there exist local coordinates $u=(u_1,\dots,u_d)$ such that
\[
K(g(u))=u^{2k},\qquad
\varphi(g(u))|g'(u)|=|u|^h b(u),
\]

with $b(u)$ positive and analytic. In other words, the theorem guarantees the existence of local coordinates in which the Kullback--Leibler function becomes monomial, or normal crossing, up to a smooth positive factor. This is the form needed to analyze both the zeta function and the marginal likelihood asymptotically; see Atiyah's classic treatment of this idea in the context of distributions~\cite{Atiyah1970}, and its use in singular learning theory in Ref.~\cite{Watanabe2009}.

Once this form is available, the local contribution to the marginal likelihood becomes an integral of monomial type, and the dominant power of $n$ is determined by the same exponents that control the poles of $\zeta(z)$~\cite{Watanabe2009,Watanabe2024Review}. In this way, the largest pole $-\lambda$ and its order $m$ govern the asymptotic expansion of the free energy.

In the present paper, however, the situation is simpler than in a generic singular model. For the aligned source-discrimination problem, the leading Kullback--Leibler functions are already of normal-crossing type in the original local coordinates, namely $D_{\mathrm{DI}}(\epsilon,s)\asymp \epsilon^2 s^2$ and $D_{\mathrm{SPADE}}(\epsilon,s)\asymp \epsilon s^2$. Therefore one does not need to carry out an additional resolution-of-singularities procedure explicitly: the zeta-function calculation can be performed directly from these local monomial forms. Resolution of singularities remains the conceptual reason that such monomial forms are the right objects to study in general, but for the present model the relevant structure is already visible without further blow-up coordinates.

\paragraph*{Relation to the present paper.}

The present source-discrimination problem is singular because the one-source null distribution is represented non-identifiably inside the closure of the two-source model. In the aligned parameterization its KL-zero set is $W_0^{\mathrm{al}}=\{\epsilon=0\}\cup\{s=0\}$. The usual regular likelihood-ratio theory is therefore not the right asymptotic framework in a neighborhood of this non-identifiable set. Instead, the free-energy asymptotics are controlled by the singular invariants $(\lambda,m)$ of the corresponding Kullback--Leibler function.

In the aligned case studied in Sec.~IV~A, this leads to

\begin{eqnarray*}
F_n^{\mathrm{DI}}&=&\frac12\log n-\log\log n+O_p(1),\\
F_n^{\mathrm{SPADE}}&=&\frac12\log n+O(1),
\end{eqnarray*}

because direct imaging and aligned SPADE have the same RLCT $\lambda=1/2$ but different multiplicities $m=2$ and $m=1$ within the smooth-positive prior class. In the fixed-offset binary-SPADE model of Sec.~IV~B, a nonzero $\theta$ removes the aligned support mismatch and gives the local normal-crossing form $K\asymp\epsilon^2s^2$; under the same prior class this yields $(\lambda,m)=(1/2,2)$. The source-position convention still changes the local order in $s$: the fixed-primary direction has $K=O(s^2)$ for fixed $\epsilon$, whereas the centroid-preserving direction generically has $K=O(s^4)$; see Appendix~B.

\section{Local zeta integrals in the aligned case}

For completeness, we record here the explicit local zeta-integral calculation underlying Eq.~\eqref{eq:zeta-poles}, and hence Eq.~\eqref{eq:RLCT-aligned}, in the aligned Gaussian case. The leading Kullback--Leibler functions are already in normal-crossing form, so no additional blow-up or resolution step is needed for the calculations below.

As explained in Sec.~IV~A and Appendix D, in a neighborhood of $(\epsilon,s)=(0,0)$, with $W_0^{\mathrm{al}}=\{\epsilon=0\}\cup\{s=0\}$, the leading Kullback--Leibler behaviors are

\begin{equation}
D_{\mathrm{DI}}(\epsilon,s)\asymp \epsilon^2 s^2,
\qquad
D_{\mathrm{SPADE}}(\epsilon,s)\asymp \epsilon s^2,
\label{eq:E1-leading-KL}
\end{equation}

up to smooth positive prefactors. Since the prior density $\phi(\epsilon,s)$ is assumed to be smooth and strictly positive near the relevant portion of $W_0^{\mathrm{al}}$, it does not alter the local vanishing orders.

Because $W_0^{\mathrm{al}}$ is the union of two components rather than an isolated point, it is useful to check the generic points of those components as well. For direct imaging, a generic point on either $\epsilon=0$ or $s=0$ has local form $K\asymp t^2$ in the transverse coordinate $t$, hence local pole data $(\lambda,m)=(1/2,1)$; the neighborhood where both factors become small has the same RLCT but multiplicity two. For aligned SPADE, a generic point on $s=0$ has $K\asymp s^2$ and hence $(1/2,1)$, whereas a generic point on $\epsilon=0$ with $s\ne0$ has $K\asymp\epsilon$ and hence $(1,1)$. Thus no omitted generic part of $W_0^{\mathrm{al}}$ produces a pole to the right of the one obtained below, and the multiplicity-two direct-imaging contribution is specific to the joint local region where both factors vanish.

Accordingly, after restricting to a sufficiently small neighborhood

\[
0<\epsilon<\delta,\qquad 0<s<\delta,
\]

the local zeta function for direct imaging has the same pole structure as

\begin{equation}
\zeta_{\mathrm{DI}}(z)
\sim
\int_0^\delta d\epsilon
\int_0^\delta ds\,
(\epsilon^2 s^2)^z
=
\left(
\int_0^\delta \epsilon^{2z}\,d\epsilon
\right)
\left(
\int_0^\delta s^{2z}\,d s
\right).
\label{eq:E2-zeta-DI}
\end{equation}

Each factor is elementary:

\begin{equation}
\int_0^\delta \epsilon^{2z}\,d\epsilon
=
\frac{\delta^{2z+1}}{2z+1},
\qquad
\int_0^\delta s^{2z}\,d s
=
\frac{\delta^{2z+1}}{2z+1},
\label{eq:E3-DI-factors}
\end{equation}

so that
\begin{equation}
\zeta_{\mathrm{DI}}(z)
\sim
\frac{C_{\mathrm{DI}}}{(2z+1)^2}
\label{eq:E4-zeta-DI-pole}
\end{equation}

for some positive constant $C_{\mathrm{DI}}$. Hence the rightmost pole is at $z=-1/2$ with multiplicity two.

For aligned SPADE, the corresponding local zeta integral has the same pole structure as
\begin{equation}
\zeta_{\mathrm{SPADE}}(z)
\sim
\int_0^\delta d\epsilon
\int_0^\delta ds\,
(\epsilon s^2)^z
=
\left(
\int_0^\delta \epsilon^{z}\,d\epsilon
\right)
\left(
\int_0^\delta s^{2z}\,d s
\right).
\label{eq:E5-zeta-SPADE}
\end{equation}

Now

\begin{equation}
\int_0^\delta \epsilon^{z}\,d\epsilon
=
\frac{\delta^{z+1}}{z+1},
\qquad
\int_0^\delta s^{2z}\,d s
=
\frac{\delta^{2z+1}}{2z+1},
\label{eq:E6-SPADE-factors}
\end{equation}

and therefore
\begin{equation}
\zeta_{\mathrm{SPADE}}(z)
\sim
\frac{C_{\mathrm{SPADE}}}{(z+1)(2z+1)}
\label{eq:E7-zeta-SPADE-pole}
\end{equation}

for some positive constant $C_{\mathrm{SPADE}}$. The rightmost pole is again at $z=-1/2$, but now with multiplicity one.

Thus,
\begin{equation}
(\lambda_{\mathrm{DI}},m_{\mathrm{DI}})
=
\left(\frac12,2\right),
\qquad
(\lambda_{\mathrm{SPADE}},m_{\mathrm{SPADE}})
=
\left(\frac12,1\right),
\label{eq:E8-rlct}
\end{equation}

which is the result quoted in Eq.~\eqref{eq:RLCT-aligned} of the main text for a prior that is smooth and strictly positive near the aligned KL-zero set $W_0^{\mathrm{al}}=\{\epsilon=0\}\cup\{s=0\}$ in the local parameter region under consideration.

\subsection{Sensitivity to boundary-weighted priors}
The preceding pole structure is not independent of the prior measure if the prior itself vanishes or diverges on the KL-zero set. To isolate this effect without introducing competing contributions from distant parts of parameter space, in this subsection we restrict the alternative to a sufficiently small rectangle $0<\epsilon<\delta$, $0<s<\delta$ and consider the fixed prior family
\begin{equation}
\phi_{a,b}(\epsilon,s)
=
C\,\epsilon^{a-1}s^{b-1}\widetilde\phi(\epsilon,s),
\qquad a,b>0,
\label{eq:E9-power-prior}
\end{equation}
where $\widetilde\phi$ is smooth and strictly positive on a neighborhood of the closure of this local rectangle and $C>0$. The smooth-positive case above corresponds to $a=b=1$. Under this explicitly local prior model, the generic components of $W_0^{\mathrm{al}}$ contribute the same candidate poles $a/2$ and $b/2$ for direct imaging, and $a$ and $b/2$ for SPADE, so the formulas below are the pole data of the complete local model--prior pair. If a global prior is allowed to have different boundary behavior on distant portions of $W_0^{\mathrm{al}}$, those portions must be checked separately for competing poles.

For direct imaging,
\begin{equation}
\zeta_{\mathrm{DI}}(z)
\sim
\frac{C_{a,b}^{\mathrm{DI}}}{(2z+a)(2z+b)},
\label{eq:E10-prior-DI}
\end{equation}
so that
\begin{equation}
\lambda_{\mathrm{DI}}=\frac12\min(a,b),\qquad
m_{\mathrm{DI}}=
\begin{cases}
2,& a=b,\\
1,& a\ne b.
\end{cases}
\label{eq:E11-prior-DI-rlct}
\end{equation}
For aligned SPADE,
\begin{equation}
\zeta_{\mathrm{SPADE}}(z)
\sim
\frac{C_{a,b}^{\mathrm{SPADE}}}{(z+a)(2z+b)},
\label{eq:E12-prior-SPADE}
\end{equation}
and therefore
\begin{equation}
\lambda_{\mathrm{SPADE}}=\min\!\left(a,\frac b2\right),\qquad
m_{\mathrm{SPADE}}=
\begin{cases}
2,& a=b/2,\\
1,& a\ne b/2.
\end{cases}
\label{eq:E13-prior-SPADE-rlct}
\end{equation}
Thus the particular equality $\lambda_{\mathrm{DI}}=\lambda_{\mathrm{SPADE}}=1/2$ with multiplicities $2$ and $1$ is a statement about the smooth-positive prior class, not about arbitrary boundary-weighted priors. More generally, Eqs.~\eqref{eq:E11-prior-DI-rlct} and \eqref{eq:E13-prior-SPADE-rlct} imply
\[
\lambda_{\mathrm{SPADE}}\ge\lambda_{\mathrm{DI}}\qquad(a,b>0).
\]
If $a>b$, the complete pole data coincide, $(\lambda,m)_{\mathrm{SPADE}}=(\lambda,m)_{\mathrm{DI}}=(b/2,1)$; if $a=b$, the RLCTs coincide but the multiplicities remain $1$ and $2$, respectively; and if $a<b$, the SPADE RLCT is strictly larger.

As a concrete family of boundary weights, take $a=1/r$ and $b=1$ with $r\ge1$. When $r\in\mathbb N$, this family is induced by writing $\epsilon=x^r$ and assigning a uniform prior to $x\in(0,1)$ while keeping a locally flat prior in $s$. The induced density in the physical brightness coordinate is
\begin{equation}
\phi(\epsilon) = \frac1r\epsilon^{1/r-1},
\label{eq:E14-induced-prior}
\end{equation}
which corresponds to $a=1/r$ and $b=1$. Hence
\begin{equation}
(\lambda_{\mathrm{DI}},m_{\mathrm{DI}})=
\begin{cases}
(1/2,2),&r=1,\\
(1/(2r),1),&r>1,
\end{cases}
\label{eq:E15-example-DI}
\end{equation}
and
\begin{equation}
(\lambda_{\mathrm{SPADE}},m_{\mathrm{SPADE}})=
\begin{cases}
(1/2,1),&r=1,\\
(1/2,1),&1<r<2,\\
(1/2,2),&r=2,\\
(1/r,1),&r>2.
\end{cases}
\label{eq:E16-example-SPADE}
\end{equation}
The character of the DI--SPADE distinction therefore changes: for $r>1$ it is no longer the flat-prior ``same RLCT, different multiplicity'' comparison and can move to the leading $\log n$ order. For this power-law family, the aligned SPADE prior-mixture exponent is not reversed relative to direct imaging; rather, the specific mechanism of the advantage changes.

It is also useful to distinguish a change of coordinates from a change of prior measure. If $\epsilon=x^r$ is treated merely as a reparameterization and the original prior measure is pushed forward consistently, the marginal likelihood and its pole data are unchanged. Taking $x$ to be flat instead defines a new prior measure concentrated differently near $\epsilon=0$, and it is this change of measure that produces Eqs.~\eqref{eq:E15-example-DI}--\eqref{eq:E16-example-SPADE}.


\begin{thebibliography}{99}

\bibitem{Tsang2016}
M.~Tsang, R.~Nair, and X.-M.~Lu,
Phys. Rev. X \textbf{6}, 031033 (2016).

\bibitem{NairTsang2016}
R.~Nair and M.~Tsang,
Phys. Rev. Lett. \textbf{117}, 190801 (2016).

\bibitem{LuKroviNairGuhaShapiro2018}
X.-M.~Lu, H.~Krovi, R.~Nair, S.~Guha, and J.~H.~Shapiro,
npj Quantum Inf. \textbf{4}, 64 (2018).

\bibitem{Paur2016}
M.~Pa\'ur, B.~Stoklasa, Z.~Hradil, L.~L.~S\'anchez-Soto, and J.~\v{R}eh\'a\v{c}ek,
Optica \textbf{3}, 1144--1147 (2016).

\bibitem{Almeida2021}
J.~O.~de Almeida, J.~Ko{\l}ody\'nski, C.~Hirche, M.~Lewenstein, and M.~Skotiniotis,
Phys. Rev. A \textbf{103}, 022406 (2021).

\bibitem{Zhang2024UnequalBrightness}
J.-D.~Zhang, Y.~Fu, L.~Hou, and S.~Wang,
Opt. Express \textbf{32}, 26147--26156 (2024).

\bibitem{Grace2020}
M.~R.~Grace, Z.~Dutton, A.~Ashok, and S.~Guha,
J. Opt. Soc. Am. A \textbf{37}, 1288--1299 (2020).

\bibitem{OzerGrace2024}
I.~Ozer, M.~R.~Grace, P.-A.~Blanche, and S.~Guha,
arXiv:2409.04323 (2024).

\bibitem{Schlichtholz2024}
K.~Schlichtholz, T.~Linowski, M.~Walschaers, N.~Treps, \L{}.~Rudnicki, and G.~Sorelli,
Optica Quantum \textbf{2}, 29--34 (2024).

\bibitem{Hartigan1985}
J.~A.~Hartigan,
in \textit{Proceedings of the Berkeley Conference in Honor of Jerzy Neyman and Jack Kiefer},
Vol.~2, pp.~807--810 (1985).

\bibitem{ChenLi2009}
J.~Chen and P.~Li,
Ann. Statist. \textbf{37}, 2523--2542 (2009).

\bibitem{Watanabe2009}
S.~Watanabe,
\textit{Algebraic Geometry and Statistical Learning Theory}
(Cambridge University Press, Cambridge, England, 2009).

\bibitem{KariyaWatanabe2022}
N.~Kariya and S.~Watanabe,
Commun. Stat. Theory Methods \textbf{51}, 5873--5888 (2022).

\bibitem{KariyaWatanabe2020}
N.~Kariya and S.~Watanabe,
IEICE Trans. Fundam. Electron. Commun. Comput. Sci. \textbf{E103.A}, 1274--1282 (2020).

\bibitem{HuangLupo2021}
Z.~Huang and C.~Lupo,
Phys. Rev. Lett. \textbf{127}, 130502 (2021).

\bibitem{Gessner2020}
M.~Gessner, C.~Fabre, and N.~Treps,
Phys. Rev. Lett. \textbf{125}, 100501 (2020).

\bibitem{Linowski2023}
T.~Linowski, K.~Schlichtholz, G.~Sorelli, M.~Gessner, M.~Walschaers, N.~Treps, and \L{}.~Rudnicki,
New J. Phys. \textbf{25}, 103050 (2023).

\bibitem{Watanabe2024Review}
S.~Watanabe,
arXiv:2406.10234 (2024).

\bibitem{Atiyah1970}
M.~F.~Atiyah,
Commun. Pure Appl. Math. \textbf{13}, 145--150 (1970).

\end{thebibliography}
\end{document}